\documentclass[twocolumn,a4paper,preprintnumbers,amsmath,amssymb,showpacs,prb]{revtex4}

\usepackage{graphicx}
\usepackage{bm}

\begin{document}


\title{Hyper-domains in exchange bias micro-stripe pattern}

\author{Katharina Theis-Br\"ohl}
\email{k.theis-broehl@rub.de}
\homepage{http://www.ep4.ruhr-uni-bochum.de}
\author{Andreas Westphalen}
\author{Hartmut Zabel}

\affiliation{Department of Physics, Ruhr-University Bochum,
D-44780 Bochum, Germany}

\author{Ulrich R\"ucker}

\affiliation{Institut f\"ur Festk\"orperforschung,
Forschungszentrum J\"ulich GmbH, D-52425 J\"ulich, Germany }

\author{Jeffrey McCord}

\affiliation{Leibniz Institute for Solid State and Materials
Research Dresden, Institute for Metallic Materials, Helmholtzstr.
20, D-01169 Dresden, Germany}

\author{Volker H\"oink}
\author{Jan Schmalhorst}
\author{G\"unther Reiss}

\affiliation{Department of Physics, University Bielefeld,
Universit\"atsstrasse 25, D-33615 Bielefeld, Germany}

\author{Tanja Weis}
\author{Dieter Engel}
\author{Arno Ehresmann}

\affiliation{Institute of Physics and Centre for Interdisciplinary
Nanostructure Science and Technology (CINSaT), University of
Kassel, Heinrich-Plett-Strasse 40, D-34132 Kassel, Germany}

\author{Boris P. Toperverg}

\affiliation{Department of Physics, Ruhr-University Bochum,
D-44780 Bochum, Germany, \\
and Petersburg Nuclear Physics Institute, Gatchina, 188300 St.
Petersburg, Russia}

\begin{abstract}
A combination of experimental techniques, e.g. vector-MOKE
magnetometry, Kerr microscopy and polarized neutron reflectometry,
was applied to study the field induced evolution of the
magnetization distribution over a periodic pattern of alternating
exchange bias stripes. The lateral structure is imprinted into a
continuous ferromagnetic/antiferromagnetic exchange-bias bi-layer
via laterally selective exposure to He-ion irradiation in an applied
field. This creates an alternating frozen-in interfacial exchange
bias field competing with the external field in the course of the
re-magnetization. It was found that in a magnetic field applied at
an angle with respect to the exchange bias axis parallel to the
stripes the re-magnetization process proceeds via a variety of
different stages. They include  coherent rotation of magnetization
towards the exchange bias axis, precipitation of small random
(ripple) domains, formation of a stripe-like alternation of the
magnetization, and development of a state in which the magnetization
forms large hyper-domains comprising a number of stripes. Each of
those magnetic states is quantitatively characterized via the
comprehensive analysis of data on specular and off-specular
polarized neutron reflectivity. The results are discussed within a
phenomenological model containing a few parameters which can readily
be controlled by designing systems with a desired configuration of
magnetic moments of micro- and nano-elements.
\end{abstract}

\maketitle

\section{Introduction}

The exchange bias (EB) effect, which is due to exchange coupling
between ferromagnetic (F) and antiferromagnetic (AF) layers, is
expressed via a shifted hysteresis loop away from zero field. The
shift is attributed to the frozen-in global unidirectional
anisotropy of the system. Due to its intriguing physics and
importance for device application the EB effect is persistently
under extensive study (see  Refs. \cite{Nougues, Berkowitz, Stamps,
Kiwi, Nogues_nano}). Spacial alteration of the EB field brings
qualitatively new physics into EB systems and creates new promising
technological perspectives. Deep understanding of the role of
competing interactions in this class of materials is required e.g.
to manufacture F/AF films with imprinted {\it in-plane}
ferromagnetic domain pattern with desired morphology. One of the
motivations is to design planar magnetic patterning of a continuous
film as an alternative to magnetic grains, clusters or structures
patterned by lithographic methods. With such systems one could avoid
the problem of very small feature sizes, where the long-term thermal
stability of the magnetic elements is lost due to superparamagnetic
fluctuations \cite{Moser}.

The EB effect is usually set via cooling the system below the
blocking temperature of the AF layer in a magnetic field saturating
the ferromagnetic counterpart. The size and the sign of the EB
effect depends on the choice of F/AF materials in contact and often
can be altered via changing the field cooling protocol \cite{Suess,
Hong, Ohldag}. On the other hand, the EB field direction and
strength can be selectively altered by ion bombardment of the F/AF
bi-layer subjected into an external field of the opposite direction
\cite{Schmalhorst, Engel}. Depositing a grid protecting some areas
of the sample one can preserve the EB field in those areas, while
altering its direction in the unprotected regions. The method of Ion
Beam Induced Magnetic Patterning (IBMP) \cite{Ehresmann_rev} opens
wide perspectives to produce various EB patterns. Here we report on
the magnetic properties of a IBMP produced  stripe-like pattern with
the EB-field set antiparallel in neighboring stripes so that the net
EB field is compensated and the system should posses global
uniaxial, instead of unidirectional, anisotropy. Therefore it is
expected that the ferromagnetic stripe domains in such a system
display alternating magnetization directions in the remanent
demagnetized state of the sample. Details on the sample preparation
and the results of experimental and theoretical studies of the
magnetization reversal mechanism of the system for the field applied
along the exchange bias axis can be found in our recent publication
\cite{Theis-Broehl_PRB2006}. There by use of the magneto-optical
Kerr-effect (MOKE) in vector-MOKE configuration, Kerr microscopy and
polarized neutron reflectometry (PNR) it has been shown that the
system exhibits a hysteresis rich in details and with a complex
domain structure. Surprisingly, it was found that in the easy axis
configuration the magnetic state after reversal of one of the both
types of stripe domains never shows a clear antiparallel domain
structure. Instead, at low fields the magnetization in the different
EB stripes is periodically canted with respect to the EB axis so
that alternating antiparallel ordering of domain magnetization
projections onto the stripe axis co-exists with a macroscopic
magnetic moment perpendicular to the anisotropy axis. This effect is
explained within the framework of a simple phenomenological model
which takes into account competing interfacial and intralayer
exchange interactions. According to the model, within a certain
range of parameters including, e.g. the ratio between ferromagnetic
layer thickness and the stripe width, the system reveals an
instability with respect to the tilt of magnetization to the left,
or to the right against the EB stripe induced anisotropy axis. In
our previous study \cite{Theis-Broehl_PRB2006} it is always found
flipped only in one of two nominally equal directions so that the
system always carried an appreciable residual magnetization not
collinear with the anisotropy axis and the magnetic field.

The finding in Ref. \cite{Theis-Broehl_PRB2006} may have more
general and far going consequences for the understanding of the
inherent physics of the EB effect. Areas with alternating EB fields
must exist in any F/AF bi-layers coupled via exchange interaction
through a common  {\em atomically rough} interface
\cite{Malozemoff_1987, Malozemoff_1988}. Alternating interfacial
fields in this case are generated by uncompensated spins in AF areas
in neighboring atomic steps which are shifted up or down with
respect to each other for  half of the  magnetic unit cell of the
Ising type antiferromagnet. Those interfacial fields randomly
alternate over the F/AF interface and compete with the lateral
exchange interaction which favors a homogeneously magnetized
ferromagnetic film.  Depending on film thickness and strength of the
interactions, e.g. the interfacial F/AF vs lateral exchange in the
ferromagnetic film, the competition may result in a state with
magnetization of the Heisenberg ferromagnet (inhomogeneously) tilted
away from the external field applied collinearly with EB direction.
In view of that the magnetization distribution in the model system
with a controlled EB field alternation deserves further
comprehensive consideration.

First of all we admit that the previously observed
\cite{Theis-Broehl_PRB2006} preferential large tilt of the net
magnetization away from the symmetry axis can be explained by a tiny
misalignment between frozen-in EB fields in irradiated and protected
stripes. Such a misalignment determines a preferential direction of
the small net EB field noncollinear with stripes. In the vicinity of
the instability point the net magnetization naturally appears in the
direction of this field. It is quite a challenging technological
task to set both EB fields exactly antiparallel and collinear with
the stripes. This is not a goal of the present paper, where instead,
we report on the results avoiding this problem by an deliberate tilt
of the external field direction by an angle as high as 45$^\circ$
regarding to the anisotropy axis set along the stripes. Then a
little misalignment between the EB fields in neighboring stripes
plays a minor role. Measurements in an asymmetric regime, on the
other hand, disclose many details on domain organization and
evolution which are otherwise hidden, but absolutely crucial for a
complete understanding of the re-magnetization mechanisms in EB
patterned arrays and other types of systems with alternating EB
fields. The bulk of information is mostly gained via the
quantitative analysis of data on PNR. The scattering intensity
distribution was measured over a broad range of angles of incidence
and scattering and accomplished with a full polarization analysis at
different magnetic fields along the hysteresis loop. For fitting the
specular reflectivity data we used an originally developed
least-squares software package, \cite{Deriglazov} which allows for
simultaneous evaluation of all four measured reflectivities in one
cycle. We compare the results of our fits  to vector-MOKE
measurements. For supporting the interpretation of our data on
off-specular diffuse and Bragg scattering Kerr-microscopy (KM)
images \cite{McCord} were also taken.

\section{Experimental Details}
The sample studied  is a TaO-Ta(8.7~nm)/
Co$_{70}$Fe$_{30}$(28.0~nm)/ Mn$_{83}$Ir$_{17}$(15.2~nm)/
Cu(28.4~nm)/ SiO$_2$(50.5~nm)/ Si(111) film stack. The initial EB
direction was set by field cooling in a magnetic in-plane field of
100~mT after an annealing step for 1 h at 275~$^\circ$C which is
above the blocking temperature of the antiferromagnetic material.
Subsequently, IBMP using He$^+$ ions with a fluency of $1\times
10^{15}$~ions/cm$^2$ at 10 keV was applied at a magnetic field of
100~mT aligned opposite to the initial EB direction. This resulted
in a stripe-like domain pattern of equally spaced stripes with a
width of $2.5~\mu$m and a periodicity of $\Lambda=5~\mu$m and with
alternate sign of the unidirectional anisotropy, and hence of the EB
in neighboring stripes.\footnote{More details on sample preparation
and treatment are given in Ref. \cite{Theis-Broehl_PRB2006}.}

The evolution of magnetization as a function of applied field was
recorded with a vector-MOKE setup as described in Ref.
\cite{Schmitte}. Both magnetization projections were accessed via
measuring the Kerr angle parallel and perpendicular to the field.
The projections were measured in the longitudinal MOKE
configuration, i.e. with the incident light linearly polarized
within the reflection plane. For determination of the magnetization
component perpendicular to the field the sample and the magnetic
field direction were rotated simultaneously by 90$^\circ$ about the
normal to the surface. Assuming that the Kerr angles,
$\theta_{\mathrm K}^{\mathrm L}\propto M\cos{\overline\gamma}$,
measuring the magnetization vector projection and the Kerr angle
parallel to the field, and $\theta_{\mathrm K}^{\mathrm T}\propto
M\sin{\overline\gamma}$, corresponding to the perpendicular
magnetization component, have the same proportionality coefficients,
one can determine the tilt angle $\overline\gamma$ through
$\tan\overline\gamma={\theta_{\mathrm K}^{\mathrm
T}}/{\theta_{\mathrm K}^{\mathrm L}}$ and  the normalized length of
the magnetization vector $M/M_{\mathrm s}={\sqrt{(\theta_K^L)^2
+(\theta_K^T)^2}}/{\theta_K^{\mathrm s}}$ with $M_{\mathrm s}$ being
the modulus of the magnetization and $\theta_K^{\mathrm s}$ the Kerr
angle, both in saturation. This allows to completely determine the
direction of the mean magnetization vector
${\bf\mbox{\boldmath$M$}}$ and its absolute value
$M=|{\bf\mbox{\boldmath$M$}}|$ reduced due to domains. In different
domains the magnetization vector deviates by angles $\Delta\gamma$
from the direction of the mean magnetization and hence is tilted by
$\gamma=\overline\gamma+\Delta\gamma$ against the applied field. The
mean angle $\overline\gamma$ is  determined by the equation
$\langle\sin\Delta\gamma\rangle_\mathrm{coh}=0$, where the averaging
runs over the spot coherently illuminated by the laser beam. Then
the normalized longitudinal and, correspondingly, transverse MOKE
signals can be written as:
\begin{eqnarray}
\langle\cos\gamma\rangle_\mathrm{coh}&=&M/M_s\cos\overline\gamma=
\langle\cos\Delta\gamma\rangle_\mathrm{coh}\cos\overline\gamma,\\
\langle\sin\gamma\rangle_\mathrm{coh}&=&M/M_s\sin\overline\gamma=
\langle\cos\Delta\gamma\rangle_\mathrm{coh}\sin\overline\gamma,
\end{eqnarray}
where the mean magnetization
$M=M_s\langle\cos\Delta\gamma\rangle_\mathrm{coh}$.

Further insight into the microscopic rearrangement of magnetization
was achieved by a high resolution magneto-optical Kerr microscope
(KM) \cite{HubertSchaefer} that is sensitive to directions
orthogonal to the field \cite{McCord}.

Neutron scattering experiments were carried out with the HADAS
reflectometer at the FRJ-2 reactor in J\"ulich, Germany. Details of
the measuring geometry can be found in Ref.
\cite{Theis-Broehl_PRB2006}. In the present experiment the sample
was aligned so that the  field orientation was tilted by $45^\circ$
against the EB axis.  The monochromatic neutron beam with the
wavelength of $0.452$~nm incident onto the sample surface under the
glancing angle $\alpha_i$ is scattered under the angle $\alpha_f$,
so that for specular reflection $\alpha_f=\alpha_i$. The incident
polarization vector ${\mbox{\boldmath $P$}}_i$ was directed either
parallel or antiparallel to the magnetic field and perpendicular to
the scattering plane. In each of the two directions of
${\mbox{\boldmath $P$}}_i$ the final spin state was analyzed with
respect to the same direction with an efficiency
$P_f=|{\mbox{\boldmath $P$}}_f|\leq1$.

Specular PNR provides information similar but not identical to
vector-MOKE magnetometry. Both methods probe projections of the
magnetization vector averaged over its deviations due to magnetic
domains and other inhomogeneities within the coherence volume of
photons, or neutrons, respectively. In the case of MOKE the laser
beam is rather coherent all over the isotropic light spot
illuminating the sample surface. This is not the case for PNR. The
neutron source is essentially incoherent, but the neutron beam is
well collimated in the reflection plane, while the collimation is
usually relaxed perpendicular to this plane. Hence the cross section
of the coherence volume with the reflecting surface is represented
by an ellipsoid with dramatically extended axis along the beam
projection onto the surface. At shallow incidence this extension,
i.e. the longitudinal coherence length, may spread up to some
fractions of a millimeter. In contrast, the other ellipsoid axis,
i.e. the coherence length across the incoming beam is short and only
amounts to about 10~nm. Therefore, the coherence 2D ellipsoid covers
only a very small area of the sample, and the observed PNR signal is
a result of two sorts over averaging. The first one includes a
coherent averaging of the reflection potential, e.g. over directions
of the magnetization in small (periodic and random) domains, and
runs over the coherence area. Second, the reflected intensities from
each of those areas are summed up incoherently.

If the mean magnetization averaged over the coherence area is
nonzero and collinear with the external field (applied parallel to
the neutron polarization axis and perpendicular to the scattering
plane) then specular reflection does not alter the neutron spin
states and only two non-spin-flip (NSF) reflection coefficients
${\mathcal R}^{+\,+}\neq {\mathcal R}^{-\,-}$ are measured, while
both spin-flip (SF) reflectivities ${\mathcal R}^{+\,-}={\mathcal
R}^{-\,+}=0$. In this case NSF reflectivities are uniquely
determined by the mean optical potential, e.g. by the mean
projection of the magnetization proportional to
$\langle\cos\Delta\gamma\rangle_\mathrm{coh}$, where $\Delta\gamma$
is the tilt angle of the magnetization in domains smaller than the
coherence length \footnote{This assumes that
$\langle\cos\Delta\gamma\rangle_\mathrm{coh}$ has the same value for
different coherence areas over the sample surface illuminated by the
neutron beam. Alternatively, additional averaging of reflection
intensity over values of the mean magnetization in different
coherence spots has to be performed.}.

If the mean magnetization direction makes an angle
$\overline{\gamma}= \langle\gamma-\Delta\gamma\rangle_\mathrm{coh}$
with the applied field then the SF reflectivities ${\mathcal
R}^{+\,-}={\mathcal R}^{-\,+}\neq0$ and are proportional to
$\sin^2\overline{\gamma}$, i.e. to the mean square of the
magnetization projection normal to the field. At the same time the
difference, ${\mathcal R}^{+\,+}-{\mathcal R}^{-\,-}$, between the
NSF reflectivities is proportional to
$\cos\overline{\gamma}$, i.e. to the projection of the mean
magnetization within the coherence volume onto the applied field
direction.  Crossing a number of small periodic and random domains
in only one direction the coherence ellipsoid still covers a very
small area of the sample. Therefore, the angle $\overline\gamma$ may
vary along the sample surface and reflectivities have to be
incoherently averaged over the spread of $\overline\gamma$. It is
important to note that NSF and SF reflectivities are complicated
nonlinear functions of
$\langle\cos\Delta\gamma\rangle_\mathrm{coh}$, which may also vary
along the sample surface. Hence, such an averaging is, in general,
not a trivial procedure.

If the value of $\langle\cos\Delta\gamma\rangle_\mathrm{coh}$ is,
however, unique \cite{Theis-Broehl_PRB2006} for the whole sample
surface, then
\begin{eqnarray}
{\mathcal R}^{+\,+}&-&{\mathcal R}^{-\,-}
\propto\langle\cos\overline{\gamma}\rangle_\mathrm{inc}\\
{\mathcal R}^{+\,-}&=&{\mathcal
R}^{-\,+}\propto\langle\sin^2\overline{\gamma}\rangle_\mathrm{inc}
\end{eqnarray}
are respectively proportional to $\cos\overline{\gamma}$ and
$\sin\overline{\gamma}$ incoherently averaged over the reflecting
surface with the proportionality coefficients nonlinearly depending
on $\langle\cos\Delta\gamma\rangle_\mathrm{coh}$. Because of
nonlinearity the parameters
$\langle\cos\overline{\gamma}\rangle_\mathrm{inc}$,
$\langle\sin^2\overline{\gamma}\rangle_\mathrm{inc}$ and
$\langle\cos\Delta\gamma\rangle_\mathrm{coh}$ can only be found via
fitting of the data for all NSF and SF reflectivities. After that
one can determine the mean value $\langle\cos\gamma\rangle\approx
\langle\cos\overline{\gamma}\rangle_\mathrm{inc}
\langle\cos\Delta\gamma\rangle_\mathrm{coh}$, under the condition:
$\langle\sin\Delta\gamma\rangle_\mathrm{coh}=0$. This constraint is
similar to that applied above for vector-MOKE and hence the mean
magnetization projection onto the field direction is in this case
equally measured by both methods: MOKE and PNR. Then agreement
between results of PNR and vector-MOKE can be used to prove the
assumption above. Alternatively, PNR is able to deliver an important
information complementing vector-MOKE data.

Due to the strong anisotropy of the coherence ellipsoid PNR can
probe a variation of $\overline\gamma$ in the corresponding
direction measuring fluctuations of the magnetization not accessible
for MOKE. In particular, with PNR one gains a direct access to the
dispersion
$\Sigma^2=\langle\sin^2\overline\gamma\rangle_\mathrm{inc}-
\langle\sin\overline\gamma\rangle_\mathrm{inc}^2\geq 0$ which
quantifies those fluctuations. If, for instance, $\Sigma=0$ these
fluctuations are absent, then at
$\langle\sin^2\overline\gamma\rangle_\mathrm{inc}=
\langle\sin\overline\gamma\rangle_\mathrm{inc}^2\neq 0$ the
homogeneous sample magnetization is homogeneously tilted by the
angle $\overline\gamma$ to the left, or to the right with respect to
the applied field. PNR is not sensitive to the sign of
$\overline\gamma$, which can be determined via vector-MOKE. On the
other hand, MOKE cannot provide any information about, e.g. the
totally demagnetized structure when
$\langle\sin\overline\gamma\rangle=\langle\cos\overline\gamma\rangle=0$.
In this case missing information can readily be retrieved from the
PNR data. This can already be seen considering two limiting
situations. The limiting value $\Sigma=1$ is achieved in the state
with large domains where the magnetization is perpendicular to the
applied field. The other limit $\Sigma=0$ is reached if
demagnetization occurs on a scale smaller than the coherence area.
In the latter case specular reflection is accompanied by
off-specular scattering.

Off-specular PNR can, in contrast to MOKE, directly measure the
spread of magnetization directions due to domains crossed with the
long axis of the coherence ellipsoid. Periodic deviations
$\Delta\gamma$ give rise to Bragg diffraction, while random
fluctuations cause diffuse scattering. The positions of the Bragg
peaks in the reciprocal space determine the period of the domain
structure along the largest coherence axis, while the extension of
diffuse scattering is due to the correlations of the random
component of the magnetization. Via fitting of intensities of
off-specular scattering one can deduce the magnetization
distribution between neighboring stripe domains
and in addition to $\langle\cos\Delta\gamma\rangle_\mathrm{coh}$ to
determine one more parameter,
$\langle\sin^2\Delta\gamma\rangle_\mathrm{coh}$, characterizing
magnetization fluctuations due to random ripple domains. Then one
can also calculate the dispersion
$\sigma^2=\langle\cos^2\Delta\gamma\rangle_\mathrm{coh}-
\langle\cos\Delta\gamma\rangle_\mathrm{coh}^2\geq0$ quantifying the
microscopic arrangement of magnetization fluctuations within the
coherence length.

Often, the set of parameters indicated above is not sufficient to
describe experimental data of PNR and to infer from it a domain
state. Such a situation takes place when the mean value
$\langle\cos\Delta\gamma\rangle_\mathrm{coh}$ averaged over the
coherence range varies along the sample surface. Then incoherent
averaging must take into account that specular reflection and
off-specular scattering occur from areas with different optical
potentials. As we shall see below, within a certain range of applied
fields the magnetization in our system evolves via formation of
large (hyper-) domains comprising a number of small stripe and
ripple domains. In contrast to the case of the conventional domain
state, now the magnetization in different hyper-domains differs not
only in directions, but also in absolute values. This is due to the
fact that each type of hyper-domains is characterized by a
particular arrangement of the magnetization over stripe and ripple
domains belonging to this type. Each of them acquires its own set of
parameters, e.g. $\langle\cos\Delta\gamma_j\rangle_\mathrm{coh}$,
$\langle\sin^2\Delta\gamma_j\rangle_\mathrm{coh}$,
$\langle\cos\overline\gamma_j\rangle_\mathrm{inc}$ and
$\langle\sin^2\overline\gamma_j\rangle_\mathrm{inc}$, where the
superscript $j$ indexes the type of hyper-domain. ${\mathcal
R}_j^{\pm\pm}$, and SF, ${\mathcal R}_j^{\pm\mp}$, reflectivities,
as well as cross sections of off-specular scattering, are also to be
indexed accordingly. Then the weights $w_j$ of different types of
hyper-domains can be determined along with the above listed
parameters via the fitting of experimental data. This would allow to
totally characterize the magnetic states of the system passing
through along the hysteresis loop.

\section{Experimental Results}
\subsection{Vector-MOKE}

\begin{figure}[tb]
    \begin{center}
    \includegraphics[clip=true,keepaspectratio=true,width=0.65\linewidth]{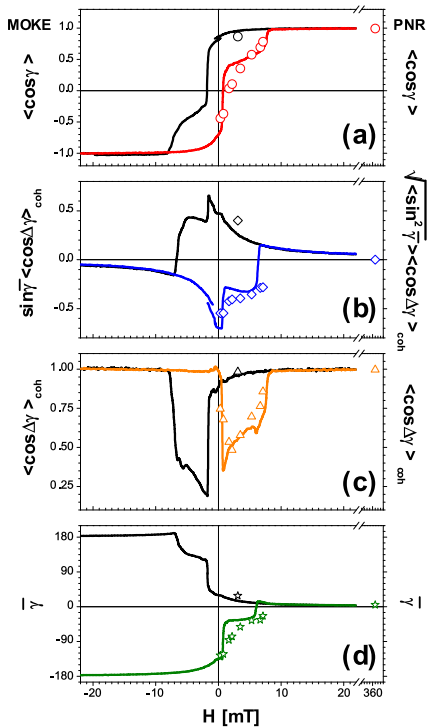}
    \caption{\label{Fig1}(Color online) Vector-MOKE data and results of fits to the PNR data.
    (a) Normalized hysteresis loop of the longitudinal magnetization component,
    (b) normalized hysteresis loop of the transverse magnetization component,
    (c) normalized value of the length of the magnetization vector, and
    (d) mean value of the angle between magnetization and external field.
    The field is applied at  $45^\circ$ with respect to the EB axis.
    Lines represent MOKE  and symbols PNR data. The symbols
    represent the results of the
    the weighted average between first and second domain (see text).
    The ascending curve is drawn in color and the descending one in black.}
    \end{center}
\end{figure}

\begin{figure}[tb]
    \begin{center}
    \includegraphics[clip=true,keepaspectratio=true,width=0.7\linewidth]{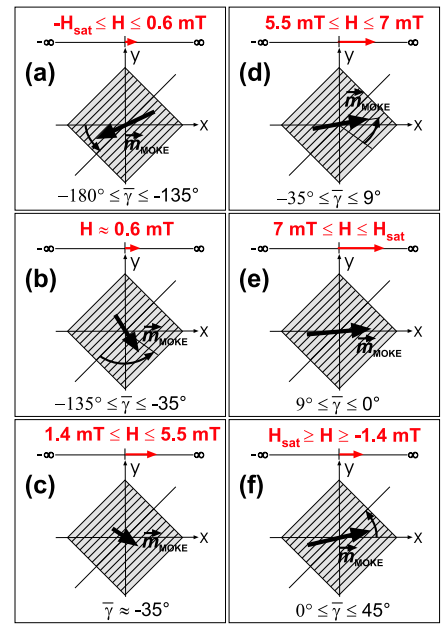}
    \caption{\label{Fig2}(Color online) Evolution of the mean magnetization along the
    MOKE hysteresis loop. }
    \end{center}
\end{figure}

The evolution of two Cartesian projections of the normalized to
saturation mean magnetization vector ${\bm m}={\bm
M}/M_\mathrm{sat}$ determined with vector-MOKE is depicted in
Fig.~\ref{Fig1}(a) and (b). Fig.~\ref{Fig1} (c) and (d) illustrate
the field dependence of the absolute value $m=\left|{\bm m}\right|$
of the normalized magnetization vector ${\bm m}$ and its tilt angle
$\overline\gamma$ with respect the field applied at the angle of
$45^\circ$ relative to the stripes axis. Comparing these plots one
can admit several stages of the re-magnetization process as
visualized in Fig.~\ref{Fig2}. Reduction of the negative field from
saturation leads firstly to relatively slow coherent rotation of the
magnetization vector away from the field direction while maintaining
its absolute value [Fig.~\ref{Fig2}~(a)]. At small positive field
$H=H_{c1}\approx0.6$ mT this process is suddenly interrupted
apparently due to domain formation [Fig.~\ref{Fig2}~(b)]. This stage
of the process is completed at $H\approx +1.4$ mT when the
magnetization loss is about  half of its magnitude. At further
increase of the applied field the magnetization partially restores
its magnitude up to almost 63\% of the nominal value. At the same
time the mean magnetization vector is directed at an angle
$\overline\gamma\approx 35^\circ$ with respect to the applied field
[Fig.~\ref{Fig2}~(c)]. Within quite a broad interval of fields the
system stays in a relatively rigid state with the mean magnetization
directed almost normal to the stripes. The next dramatic event
occurs at $H_c2$ between $H\approx5.5$ mT and $H\approx7.0$ mT when
the magnetization again looses and partially restores its absolute
value [Fig.~\ref{Fig2}~(d)]. Within this stage the vector ${\bm m}$
rapidly starts to rotate towards the magnetic field direction and is
aligned along the field at $H\approx6.0$ mT. At higher fields the
magnetization, surprisingly, continues to rotate further away from
the field direction and at $H\approx6.5$ mT it arrives at a maximal
tilt angle $\overline\gamma\approx 9^\circ$. This stage of the the
re-magnetization process is apparently governed by a domain
re-arrangement mechanism which restores the magnetization absolute
value up to about 65\% of its nominal. Rapid restoration of
magnetization follows, presumably, through an additional
intermediate stage of domain evolution within the field frame
$7.0\leq H\leq H_{sat}$ [Fig.~\ref{Fig2}~(e)].

The descending branch of the hysteresis loop [the first stage is
shown in Fig.~\ref{Fig2}~(f)]  generally repeats all main steps of
the magnetization evolution. However, the magnetization vector does
not passes them in the exactly reversed order, as would be expected.
Instead, along the descending branch the vector ${\bm M}$  continues
to rotate in the same direction passing by the state with the
magnetization along the field as in the case of the ascending branch
of the hysteresis loop. Finally the vector ${\bm m}$ accomplishes
the full circle of $360^\circ$ and then slightly rotates back
approaching negative saturation. The intrinsic reason of such a
behavior should find its explanation below after more detailed
analysis of the complete scope of the data.

Here we just mention that the hysteresis loops are shifted
exhibiting a global EB effect and hence a residual frozen-in
magnetic field. This indicates an incomplete compensation of fields
frozen-in different sets of stripes. Further insight into the
arrangement of magnetization over stripes is gained by use of Kerr
microscopy.

\subsection{Kerr microscopy}
\begin{figure}[tb]
    \begin{center}
    \includegraphics[clip=true,keepaspectratio=true,width=0.95\linewidth]{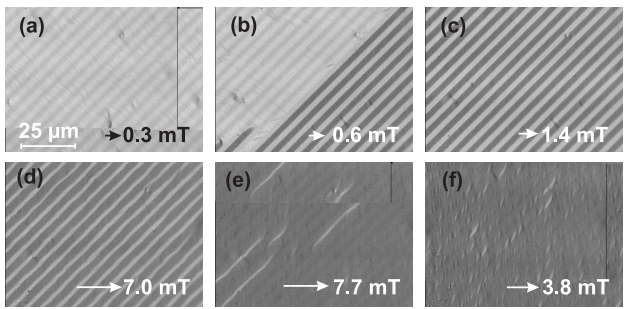}
    \caption{\label{Fig3}
    Kerr microscopy images taken  during the magnetization reversal with the field aligned diagonally to the EB axis.
    The
    images were recorded with a mixed Kerr sensitivity tuned nearly to the
    transverse magnetization direction.
    }
    \end{center}
\end{figure}

A sequence of KM images  taken along the hysteresis loop for the
present sample with the field applied along the exchange bias axis,
perpendicular, and at $45^\circ$ to this axis were briefly reported
recently \cite{McCord}. It was demonstrated, that in the latter case
the evolution of the magnetization distribution recorded in the
images, e.g. in those presented in Fig.~\ref{Fig3}, elucidates the
role of various mechanisms involved in the re-magnetization process
according to the typical stages of the re-magnetization process
shown in Figs.~\ref{Fig2} (a-f). Figs.~\ref{Fig3}(a-e) illustrate
the re-magnetization scenario along the ascending branch after
saturation in a negative field. Fig.~\ref{Fig3}(a) shows a
measurement performed at a small positive field of 0.3~mT. At this
field the magnetization already appreciably deviates from the field
axis but it is not yet reversed in neither of the stripes. The
periodic KM contrast indicates an angle between the directions of
magnetization in neighboring stripes. Some rippling, predominantly
in the He$^+$ bombarded regions, can also be observed. At 0.6~mT
[Fig.~\ref{Fig3}(b)] the reversal occurs through the formation of
large hyper-domains separated along one of the stripes. In one of
such hyper-domains the magnetization of the stripes is not yet
reversed. In the other hyper-domain the magnetization in one set of
stripes, i.e. in this case in the He$^+$ bombarded, is reversed as
seen via a large optical contrast. At 1.4~mT no hyper-domains are
seen and the magnetization projections onto the EB axis in
neighboring stripes seems to be aligned antiparallel
[Fig.~\ref{Fig3} (c)]. Further increase of the applied field changes
the scenario of the re-magnetization process, as seen in
Figs.~\ref{Fig3} (d) and (e). Now it proceeds through gradual
decrease of the width of stripes with the most unfavorable
magnetization direction. Fig.~\ref{Fig3} (f) was taken in the
backward branch. It shows that reappearance of the continuous
stripes with a negative projection onto the field direction is
preceded by precipitation of small ripple domains. Further decrease
of the positive field and its subsequent alternation restores the
periodic structure via coalescence of ripple domains in
corresponding sets of stripes \cite{McCord}. This process, however,
does not occur simultaneously all over the sample surface, but again
involves the formation of hyper-domains. Some of them contain
homogeneously magnetized stripes, while in the others the
magnetization of one set of stripes is broken into ripple domains.

It should be noted that KM images sample, in contrast to vector-MOKE
and PNR, quite a restricted area of the surface. They provide,
however, a rather solid background necessary for a quantitative
analysis of PNR data and consequent characterization of
magnetization distribution between stripe, ripple and hyper-domains
over the whole sample.

\subsection{Neutron scattering}
Neutron scattering experiments were performed  using  a
position sensitive detector (PSD). The PSD records, additionally to
the specular reflection from the mean neutron optical potential,
magnetic Bragg diffraction from the periodic stripe array and
off-specular diffuse scattering from domains smaller than the long
axis of the coherence ellipsoid. The data were taken at several
positive field values of the hysteresis. The magnetic field  was
kept parallel to the field guiding the neutron polarization in order
to avoid neutron depolarization. Prior to the measurements, the
sample was saturated in a negative field. Specular reflectivities
were extracted from the PSD maps. In Fig.~\ref{Fig4}, several
representative  experimental curves together with fits to the data
are displayed. Most of the presented data are collected at fields
corresponding to the KM images in Fig.~\ref{Fig3}. This allows for a
qualitative interpretation of the specular reflectivity curves along
with the KM measurements in Fig.~{\ref{Fig3}.

\begin{figure}
\begin{center}
    \includegraphics[clip=true,keepaspectratio=true,width=0.8\linewidth]{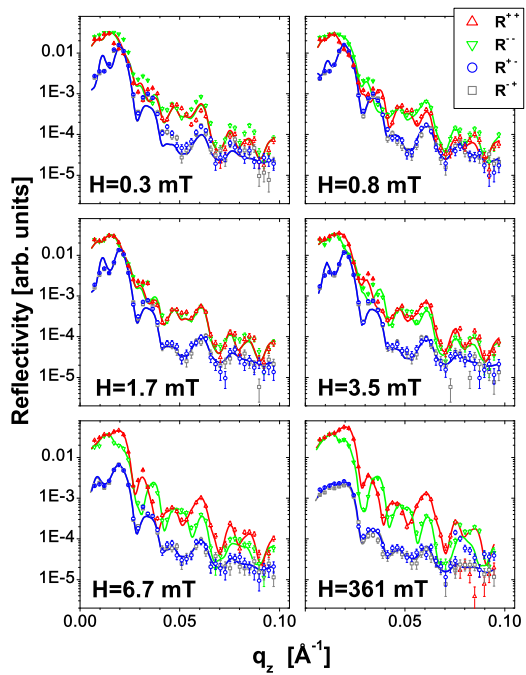}
    \caption{\label{Fig4}(Color online) Polarized neutron
reflectivity measurements performed at different magnetic fields
applied at the angle $\chi=45^\circ$ with respect to the EB axis.
The symbols present measurements of non-spin flip reflectivities
${\mathcal R}^{+\,+}$ and ${\mathcal R}^{-\,-}$ and spin flip
reflectivities ${\mathcal R}^{+\,-}$ and ${\mathcal R}^{-\,+}$. The
lines represent fits to the data points. }
    \end{center}
\end{figure}

At 0.3~mT splitting of the NSF reflectivities with ${\mathcal
R}^{-\,-}$ being higher than ${\mathcal R}^{+\,+}$ and considerable
SF reflectivities indicate an appreciable tilt of the net
magnetization, almost homogeneous in accordance with Fig.~\ref{Fig3}
(a), away from the direction antiparallel to the applied field.

At 0.8~mT the SF reflectivity is slightly increased and the
splitting of the NSF reflectivities is reduced compared to 0.3~mT.
This is attributed to a further increasing tilt of magnetization and
reduction of its absolute value due to stripe domains seen in
Fig.~\ref{Fig3} (b).

At 1.7~mT the splitting of the NSF reflectivities almost vanishes.
This qualitatively can be explained by a large angle between
magnetization directions in neighboring stripes. At the same time
the SF reflectivity attains a maximum value manifesting a large
projection of the mean magnetization component normal to the field
as seen in Fig.~\ref{Fig1}. Hence, already qualitative analysis of
the specular PNR and MOKE data let us conclude that the
magnetization vectors in neighboring stripes in Fig.~\ref{Fig3} (c)
are not collinear with either the magnetic field or the EB axis and
make quite a large angle between themselves. This angle can
precisely be determined via the quantitative analysis of the
complete scope of the PNR data.

We have undertaken further PNR measurements at 2.2~mT  (not shown
here) and 3.5~mT. At 3.5~mT increased splitting of the NSF and
reduced SF reflectivities may indicate that the mean magnetization
now turns towards the direction of the applied field. This
conclusion, however, seems to contradict the MOKE data in
Fig.~\ref{Fig1} which do not show a substantial rotation of the mean
magnetization within this interval of fields. Subsequent
quantitative PNR analysis removes the contradiction between the
vector-MOKE observations and the PNR analysis based on intuitive
arguments.

Further PNR measurements at 5.3~mT (not shown), 6.7~mT and 7.1~mT
(not shown) exhibit a continuous increase of the splitting of the
NSF and a reduction of the SF reflectivities.  The situations at
6.7~mT and 7.1~mT are comparable with the KM measurement in
Fig.~{\ref{Fig3} (d) showing gradual shrinking and final collapse of
the set of stripes with unfavorable magnetization. The SF
reflectivities are already smaller as compared to the situation
before the first reversal at 0.3~mT. Taking a closer look, one can
also admit a number of particular details distinguished in different
plots for PNR, and in particular, those recorded at 0.3 and 7.1 mT.
It is rather difficult to guess a physical meaning for most of the
changes in the PNR $q-$dependencies. Nonetheless, the least square
routine, as we shall see, allows us to infer a variation of a few
field dependent parameters quantifying KM and MOKE observations.

The major part of irrelevant parameters, e.g. those independent of
the  applied field are fixed via fitting the data collected at
361~mT, assuming the system at this field is in saturation. The
maximum splitting seen in the last plot in Fig.~{\ref{Fig4} for the
NSF reflectivities indicates that the magnetization is aligned along
the applied field. Little SF reflectivity is observed due to a not
perfect efficiency of the polarization device and taken into account
in the subsequently applied least square routine. We also performed
a measurement at a positive field of 3.1~mT in the backwards branch.
It still shows a strong splitting of NSF but an increased SF
reflectivity compared to saturation indicating a tilted
magnetization.

\begin{figure}
    \begin{center}
    \includegraphics[clip=true,keepaspectratio=true,width=0.9\linewidth]{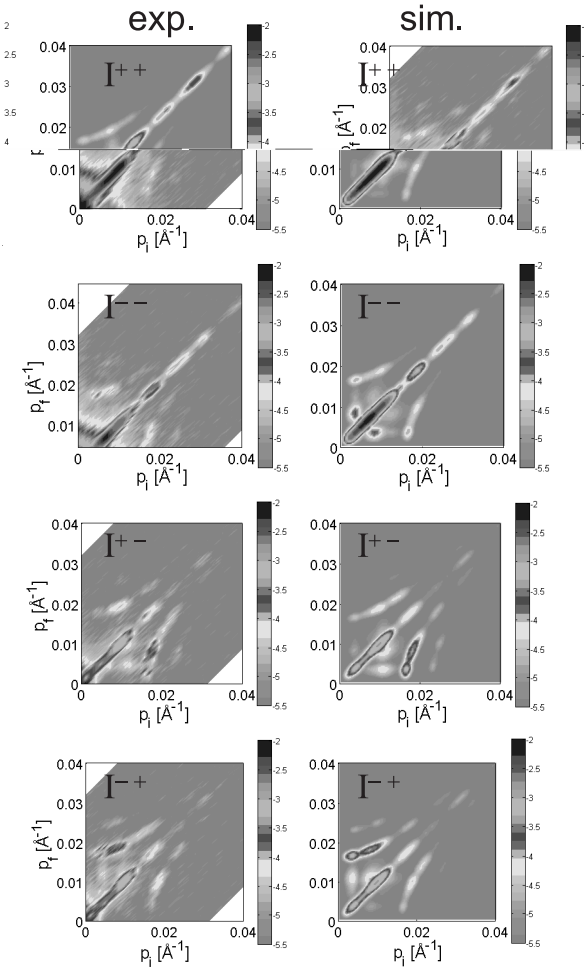}
    \caption{\label{Fig5}(Color
online) Experimental (left column) maps for the polarized neutron
scattering intensity on a logarithmic scale, measured at  a
magnetic fields of 6.7~mT. The intensities of the $I^{+\,+}$,
$I^{-\,-}$ non-spin-flip, and the intensities of the $I^{+\,-}$,
$I^{-\,+}$ spin-flip cross-sections are plotted as a function of
the angle of incidence $\alpha _i$, and the scattering angle
$\alpha_f$. Left column represents respective maps calculated in
DWBA. }
    \end{center}
\end{figure}

The most complete information on the microscopic arrangement of
magnetization is, however, obtained by analyzing not solely the
specular reflection, but in accordance with off-specular scattering.
Figure~\ref{Fig5} displays experimental data (left column) along
with the results of theoretical simulations (right column) for all
four scattering cross sections $I^{+\,+}$, $I^{-\,-}$, $I^{+\,-}$,
and $I^{-\,+}$ collected into a set of maps. The scattering
intensities are plotted as functions of the normal to the surface
components $p_i=k\sin\alpha_i$ and $p_f=k\sin\alpha_f$ of,
correspondingly, incoming, ${\bm k}_i$, and outgoing, ${\bm k}_f$,
wave vectors impinging onto the surface at glancing angles of
incidence, $\alpha_i$, and scattered at angles $\alpha_f$. The maps
were obtained for all fields listed in Fig.~\ref{Fig4} and in the
text above but here we present only those constructed for one field
at 6.7~mT and containing all features significant for the subsequent
quantitative analysis. In the maps the specular reflection ridge
runs along the diagonal, where $p_i=p_f$. At $p_i\neq p_f$ the
scattering maps exhibit two other remarkable features. The first one
is the intensity of Bragg diffraction concentrated along curved
lines
$q_\parallel\approx (2\pi n/\Lambda) \cos\chi$, where
$q_\parallel=(2\pi/\lambda)(\cos\alpha_i-\cos\alpha_f)$, $n$
denotes the order of diffraction, $\Lambda$ is the period, and
$\chi$ is the angle between the stripe axis and the normal to the
reflection plane \footnote{At $\chi=0$ the stripes run
perpendicular to the specular reflection plane and
$q_\parallel\approx (2\pi n/\Lambda)$. It should be noticed that by
rotating the sample by an angle $\chi\neq0$ Bragg reflections can
be observed \cite{Theis-Broehl_PRB2003,{toperverg00}} at smaller
angles until they merge to the reflection ridge at
$\chi\rightarrow\pi/2$.}. Bragg scattering occurs due to the
periodic variation of the magnetization across the striped pattern
and, in particular, due to the periodic alteration of the sign of
$\langle\cos\Delta\gamma\rangle_\mathrm{coh}$ in neighboring
stripes. The diffracted intensity vanishes in saturation (not shown
here) and should reach maximum values in the field range where
antiparallel alignment of the magnetization in neighboring
microstripes is expected. The second feature is  well-structured
intensity of diffuse scattering observed at low angles of incidence
$\alpha_i$ and/or scattering  $\alpha_f$. Both features are due to
the lateral magnetization fluctuations on a scale smaller than the
coherence range.

In Fig.~\ref{Fig5}, strong Bragg reflections for $n=\pm 1$ and
weaker ones for $n=\pm 2$ can be recognized. The observation of
second-order Bragg reflections is quite a striking result. In the
case of perfect alternation of magnetization projections in
neighboring stripes of equal widths Bragg reflections of all even
orders should be heavily suppressed due to the AF structure factor.
Hence reflection of the second order was never observed in our
previous measurements \cite{Theis-Broehl_PRB2006} carried out at
$\chi=0$. In the present case of $\chi=45^\circ$ one of the KM
images in Fig.~\ref{Fig5}, e.g. at $H\approx7.0$ mT, clearly
indicates a difference in the widths of neighboring stripes. This
difference violates the cancellation law for the Bragg reflection
of the even order \footnote{Higher order Bragg reflections are
suppressed by the stripe form-factor and are not detectable in
either the former \cite{Theis-Broehl_PRB2006}, or the present
experiment.}. Interestingly, second-order reflections can be
observed not only when they are expected from the corresponding KM
image in Fig.~\ref{Fig5}, but rather at all fields along the
ascending branch below saturation. In view of this, one should
admit that the cancellation law requires a perfect reciprocity
between magnetic moments of neighboring stripes. It can be violated
not only because of a difference in the stripe widths, but also due
to non-perfect alternation of stripe magnetization projections.
This is particularly the case if the magnetic field is applied at
an angle $\chi\neq0$ with respect to the main symmetry axis. Then,
in contrast to the symmetric case $\chi=0$,
\cite{Theis-Broehl_PRB2006} the external field tilts the
magnetization vector by the angle $\Delta\gamma=\beta_1$ in one set
of stripes, or by $\Delta\gamma=\beta_2$ in the other set. In the
asymmetric case $\chi\neq0$ there is no reason to expect that
$\beta_1=-\beta_2$, while at $\beta_1\neq-\beta_2$ neither of the
stripe magnetization projections perfectly alternate.

Bragg reflections are observed in all four, SF and NSF, channels.
The SF maps show a strong asymmetry with respect to the interchange
$p_i$ with $p_f$ corresponding to parity between Bragg reflections
with indexes $n$ and $-n$. The asymmetry is explained by the
birefringence \cite{toperverg99} in the mean optical potential and
is accounted for within the framework of the distorted wave Born
approximation (DWBA) \cite{toperverg02,zabel07}.  In the symmetric
case ($\chi=0$) $\beta_1=-\beta_2$ and only the magnetization vector
components collinear with the stripes alternate
\cite{Theis-Broehl_PRB2006}. Then the SF diffraction is a result of
the superposition of two effects: Bragg diffraction due to
periodical alternation of the scattering potential and a homogeneous
transverse magnetization which mixes up neutron spin states. This
superposition is described in DWBA. In the present arrangement
$\chi\neq0$ both in-plane projections of the stripe magnetization
vectors alternate, providing either NSF and SF Bragg diffraction
already in the Born approximation. However, an accurate balance
between intensities in all channels as well as between specular and
Bragg diffraction is only possible to account for accurately in
DWBA.

In Fig.~\ref{Fig5} we also observe diffuse off-specular scattering,
which is due to random fluctuations $\Delta\beta_1$ and
$\Delta\beta_2$ around their mean values $\overline\beta_1$ and
$\overline\beta_2$, respectively. The SF diffuse scattering  is also
strongly asymmetric and the asymmetry degree depends on the net
magnetization projection onto the field guiding neutron
polarization. In the $I^{+\,-}$ map off-specular scattering
intensity is mostly disposed at
$p_i<p_f$, while in the $I^{-\,+}$ one it is concentrated at
$p_i>p_f$. NSF diffuse scattering is, in contrast, symmetric. SF and
NSF diffuse scattering together indicate that there are
magnetization fluctuations of both, longitudinal and transverse
components. Off-specular diffuse scattering is strongly connected to
the development of ripple domains and is most pronounced just below
the first (at 0.3~mT) and around the second coercive field (at
7.1~mT). At fields between the both coercive fields the diffuse SF
intensity is much lower with minimum values at 3.5~mT, accounting
for a much more regular domain state. Interestingly, in the
descending branch at 3.1~mT we observe strongest diffuse SF
scattering and no Bragg reflections. The reason for such behavior is
nicely visualized in the KM measurement in Fig.~\ref{Fig3}(e) with
strong ripple development and almost no stripe contrast accounting
for a similar magnetization orientation in both stripe regions.

\section{Data analysis and discussion}

As it has already been mentioned the PNR data, although containing
a bulk of information, require a theoretical model for their
quantitative interpretation. Such a model founded on the
vector-MOKE results and in particular on the KM images which imply
the existence of at least two types of hyper-domains comprising a
number of EB stripes. This means that the net magnetization vector
\begin{equation}
{\bm M}=w_1\langle{\bm M}_1\rangle+w_2\langle{\bm M}_2\rangle
\end{equation}
\begin{figure}
    \begin{center}
    \includegraphics[clip=true,keepaspectratio=true,width=0.8\linewidth]{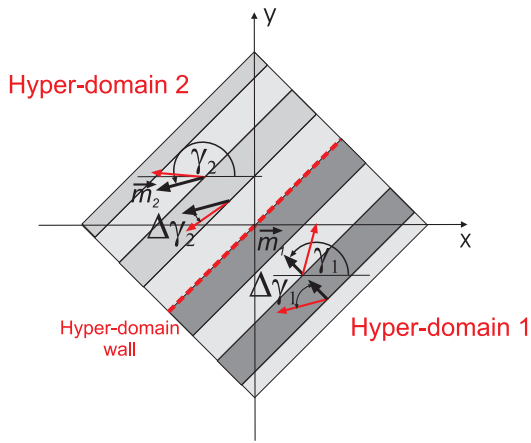}
    \caption{\label{Fig6}(Color online) Sketch of two hyper-domains separated
    by
    the domain wall (red dashed line) along one of the stripe boundary.
    The magnetization vectors of the hyper-domains are denoted by black arrows tilted
    by the angles $\gamma_{1,2}$ against the applied field. The red arrows indicate
    the local magnetization vector tilted randomly and/or periodically by angles
    $\Delta\gamma_{1,2}$ with respect to the hyper-domain magnetization directions}
    \end{center}
\end{figure}
is the sum of hyper-domain magnetization vectors $\langle{\bm
M}_1\rangle$ and $\langle{\bm M}_2\rangle$ (see, Fig.~\ref{Fig6})
weighted in accordance with the percentages $w_1$ and $w_2=1-w_1$
of the sample area they cover. The directions of the vectors of the
local magnetization ${\bm M}_{1,2}=\langle{\bm
M}_{1,2}\rangle+\Delta{\bm M}_{1,2}$ may vary as a function of the
lateral coordinates $x, y$ so that $\Delta{\bm M}_{1,2}$ describes
local deviations of the local magnetization from their mean values
$\langle{\bm M}_{1,2}\rangle$ averaged over each of the
hyper-domains. Due to the strong anisotropy in the neutron
coherency the averaging of the PNR signal, as it was pointed out
above, should be performed in two steps. Firstly, the magnetization
vectors are averaged over the coherence ellipsoids which are
dramatically extended along one axis but still are fitting into any
of the hyper-domains. In the particular kinematics the long axis is
parallel to the $x-$axis and the coherent averaging results in
$\langle{\bm M}_{1,2}(x,y)\rangle_\mathrm{coh}=\overline{\bm
M}_{1,2}(y)$. The absolute values $\left|\overline{\bm
M}_{1,2}(y)\right|$ of these vectors determine the magnetic parts
of the optical potentials and reflection amplitudes
$R^\pm_{1,2}(y)$ for each type of hyper-domain. The amplitudes may
still vary as a function of the $y$-coordinate and the equations
for SF and NSF reflection intensities require secondly an
additional incoherent averaging of the corresponding cross sections
along the $y$-direction. Those two types of averaging give access
to not only the mean values $\langle{\bm
M}_{1,2}\rangle=\langle\overline{\bm
M}_{1,2}(y)\rangle_\mathrm{inc}$ in Eq.(5) and, finally, to the net
magnetization vector ${\bm M}$, but also to the weights $w_{1,2}$.
Moreover, the least square routine provides us with quite a few
parameters rather characterizing the domain model in great details.

Our model assumes that the projections
$M^x_{1,2}(x,y)=M_\mathrm{sat}\cos\gamma_{1,2}$ and
$M^y_{1,2}(x,y)=M_\mathrm{sat}\sin\gamma_{1,2}$ of the magnetization
vectors are determined by the angles
$\gamma_{1,2}=\gamma_{1,2}(x,y)=\overline{\gamma}_{1,2}(y)+\Delta\gamma_{1,2}(x,y)$,
where $\Delta\gamma_{1,2}(x,y)$ are  describing deviations of the
angles in directions of the vectors ${\bm M}_{1,2}(x,y)$ from that
of $\overline{\bm M}_{1,2}(y)$. The latter are tilted by
$\overline{\gamma}_{1,2}(y)$ against the $x-$axis and are determined
by the constrains
$\langle\sin\Delta\gamma_{1,2}\rangle_\mathrm{coh}=0$ specific for
each type of hyper-domains and the $y$-coordinate. The deviation in
angles $\Delta\gamma_{1,2}$, either random (ripple domains) and/or
periodic (stripes), reduce the absolute values $\left|\overline{\bm
M}_{1,2}(y)\right|=M_\mathrm{sat}c_{1,2}(y)$ by the factors
$c_{1,2}(y)=\langle\cos(\Delta\gamma_{1,2})\rangle_\mathrm{coh}\leq1$.
These factors generally depend on the $y$-coordinate. However, if
the coherence length crosses a large number of stripes and/or ripple
domains this dependency is weak and can be neglected in the first
approximation so that only two parameters $c_1$ and $c_2$
characterizing reflection amplitudes are used in the fitting
routine. Two other couples of parameters
$\overline{C}_{1,2}=\langle\cos\overline\gamma_{1,2}(y)\rangle_\mathrm{inc}$
and
$\overline{S^2}_{1,2}=\langle\sin^2\overline\gamma_{1,2}(y)\rangle_\mathrm{inc}$
used to fit the data follow from the incoherent averaging of the PNR
cross sections. Such an averaging accounts for fluctuations of the
angles $\overline\gamma_{1,2}(y)$ determined for different coherence
ellipsoids. These fluctuations can be rather developed due to e.g.
ripple domains which size is greater than the coherence length in
the $y$-direction.

The set of equations for NSF and SF reflectivities used in the data
fitting are written as follows \cite{toperverg02,zabel07} :
\begin{eqnarray}
{\mathcal R}(P_i,P_f)&=w_1{\mathcal R}_1(P_i,P_f)+w_2{\mathcal
R}_2(P_i,P_f)&\nonumber\\
{\mathcal R}_{1,2}(P_i,P_f)= \frac{1}{4}\{
&[|R^+_{1,2}|^2+|R^-_{1,2}|^2][1+P_iP_f\overline{C^2}_{1,2}]&
\nonumber\\
+&[|R^+_{1,2}|^2-|R^-_{1,2}|^2](P_i+P_f)\overline{C}_{1,2}&
\nonumber\\
&+2\Re(R^{+*}_{1,2}R^-_{1,2})P_iP_f\overline{S^2}_{1,2}\},&
\label{eq:2}
\end{eqnarray}
where $P_i=\pm|P_i|$, $P_i=\pm|P_i|$ with $|P_i|\leq1$ and
$|P_i|\leq1$ are efficiencies of the polarizer and analyzer,
respectively. The complex reflection amplitudes $R^\pm_{1,2}=R(\pm
c_{1,2})$ are determined for two parameters $c_1$ and $c_2$.

At each value of applied field the best fit of all 4 measured
reflection curves was obtained by varying 7 essential parameters:
$c_{1,2}$, $\overline{C}_{1,2}$,
$\overline{S^2}_{1,2}=1-\overline{C^2}_{1,2}$ and $w_{2}=1-w_1$
while keeping all others found from the fit at saturation where
$c_{1,2}=1$, $\overline{C}_{1,2}=1$, $\overline{S^2}_{1,2}=0$ and
$w_1=1$. The quality of the fit is illustrated in Fig.~{\ref{Fig4},
while the results are collected into Fig.~{\ref{Fig7}, where the
field variation of the parameters is presented.
\begin{figure}[tb]
    \begin{center}
    \includegraphics[clip=true,keepaspectratio=true,width=0.6\linewidth]{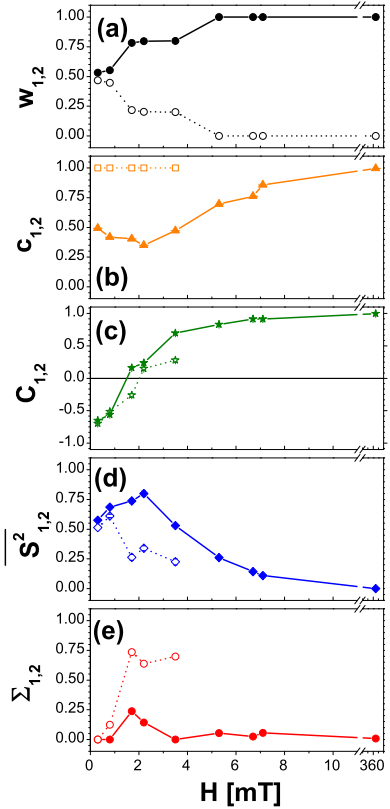}
    \caption{\label{Fig7}(Color online) Results of fits to the PNR data with respect to
        the two hyper-domains and as function of the magnetic field.
    (a) Fraction $w_1$ of hyper-domain 1,
    (b) reduction factor $c_1=\langle\cos(\Delta\gamma_{1})\rangle_\mathrm{coh}$,
    (c) parameters
    $\overline{C}_{1,2}=\langle\cos\overline\gamma_{1,2}(y)\rangle_\mathrm{inc}$,
    (d)
    $\overline{S^2}_{1,2}=\langle\sin^2\overline\gamma_{1,2}(y)\rangle_\mathrm{inc}$,  and
    (e) dispersion $\Sigma_{1,2}=\overline{C^2}_{1,2}-\overline{C}_{1,2}^2$.
   The solid symbols
    represent the results of fits to the PNR of the first hyper-domain and the hollow symbols that of the second hyper-domain.
     Lines in (b)-(d) represent respective results of the MOKE  measurements for comparison. Please note that in (d)
     $\overline{S^2}$ can only be achieved from PNR.  Instead, the PNR data are compared to $\overline{S}^2$ as calculated from $\overline{S}$ from MOKE.
    In (e) the lines are guides to the eye.}
    \end{center}
\end{figure}

First of all one can admit that two types (Fig.~{\ref{Fig7}(a)) of
hyper-domains, one with reduced (Fig.~\ref{Fig7}(b)) and the other
with saturation magnetization, exist almost all over the range of
the hysteresis loop. At low fields $w_1\approx w_2\approx 0.5$, i.e.
the two domains states are equally populated. An increasing field
$H$ leads to a two step grow of the fraction $w_1(H)$ of
hyper-domains with reduced magnetization on the cost of those
saturated until $w_1=1$ and $w_2=0$ is reached just below the second
coercive field. The reduction factor $c_1(H)$  of magnetization in
the first type of hyper-domains plotted in Fig.~{\ref{Fig7}(b) is
mostly attributed to the angle between the magnetization vectors in
neighboring stripes. It is not a monotonous function of  field $H$
and $c_1(H)$ has a minimum at $H\approx $ 2.2~mT where this angle
has a maximum value. In accordance to Fig.~{\ref{Fig7}(c) the
magnetization vectors of both types of hyper-domains rotate towards
the magnetic field direction but with a different rate. The
magnetization of "striped" hyper-domains approaches the applied
field direction much faster than that of hyper-domains with its
magnetization close to saturation. Note that at the two lowest
values $H=0.3$ and $H=0.8$~mT the magnetization of both types of
domains has a negative projection onto the field direction. At the
next measured field $H=1.7$~mT the parameter $\overline{C}_1>0$,
meaning that the magnetization vector $x$-projection of the
unsaturated hyper-domains now is  positive. At the same field
$\overline{C}_2<0$, i.e. the magnetization vectors of the saturated
hyper-domains is tilted with respect to the field by  angles of
$90^\circ<\overline\gamma_2<270^\circ$. Further increase of magnetic
field pull both magnetization vectors to the field direction.

Eq.(6) contains only
$\overline{S^2}_{1,2}=\langle\sin^2\overline\gamma\rangle$ plotted
in Fig.~{\ref{Fig7}(d), but not
$\langle\sin\overline\gamma\rangle^2$. Therefore, as already
mentioned, with PNR alone we cannot  determine the sense of
rotation. The missing information is compensated due to the
vector-MOKE measurements presented in Fig.~{\ref{Fig1}, while PNR
provides access to the dispersions
$\Sigma_{1,2}=\overline{C^2}_{1,2}-\overline{C}_{1,2}^2$, additional
physical parameters plotted in Fig.~{\ref{Fig7}(e). These quantities
measure a degree of magnetization vector fluctuations in
hyper-domains of the same type. Hence, $\Sigma_1\ll1$ all over the
range of fields signifying on rather coherent rotation of the
magnetization vector in different hyper-domains with reduced
magnetization. For the other type of hyper-domains $\Sigma_2$, on
the contrary, is rather high and reveals non-monotonous behavior.
This means that the magnetization vector in different highly
saturated hyper-domains is tilted by quite different angles
$\overline\gamma_2$ indicating the fact that there actually exist
more types of hyper-domains with nearly saturated magnetization.
Such hyper-domains can be distinguished on some of the KM images
\cite{McCord}, but unfortunately, the quality of our present PNR
data seems not allowing to introduce more parameters for their
identification.

However, our data, e.g. presented in Fig.~{\ref{Fig5} are well
sufficient to infer the microscopic magnetic structure of
hyper-domains and prove the model sketched in Fig.~{\ref{Fig6}.
Indeed, using parameters found from the fit of specular
reflectivities we can identify the reference state of the system
perturbed by periodic (stripe domains) and random (ripple domains)
deviations of magnetization from its mean value in each type of
hyper domains. Comparing intensity along the diffraction lines in
the maps in Fig.~{\ref{Fig5} with specularly reflected intensity we
determined the angles between magnetization directions in stripe
domains and direction of magnetization of hyper-domain. For
instance, at 6.7~mT it was found that $\overline\beta_1=30^\circ$
and $\overline\beta_2=0^\circ$, i.e. the angle between magnetization
directions in neighboring stripes amounts $30^\circ$. The tilt
angles was, actually, found by taking into account, that the
intensity of the Bragg lines is reduced due to ripple domains. They
cause magnetization fluctuations reducing the magnetic scattering
contrast between stripe domains and simultaneously creating diffuse
scattering seen in Fig.~{\ref{Fig5}. The intensity of diffuse
scattering allows us to quantify the amplitude of those
fluctuations. In the case of $H=$6.7~mT ripple domains reduce stripe
magnetization for about 10\% with respect to saturation. However,
fluctuations are not absolutely random but correlated over a few of
neighboring stripes as is found from the extension of diffuse
scattering in Fig.~{\ref{Fig5}. The role of correlated magnetization
fluctuations \cite{Theis-Broehl05} is not yet very clear, but they
certainly strongly influence a formation of hyper-domains in the
alternating EB systems and hence a scenario of the re-magnetization
process.

\section{Summary}

In summary, by using   vector-MOKE magnetometry, Kerr microscopy and
polarized neutron reflectometry we studied the field induced
evolution of the magnetization distribution of a periodic pattern of
alternating exchange bias stripes when applying the magnetic field
at an angle of 45$^\circ$ in order to avoid the instability with
respect to the tilt of magnetization found for the easy axis
configuration \cite{Theis-Broehl_PRB2006}. The data show that the
re-magnetization process  proceeds through different stages which
are qualitatively described by comprehensive analysis of specular
and off-specular PNR data. Beside the formation of stripe-domains
with alternated in-plane magnetization at small fields  small ripple
domains and two types large hyper-domains develop comprising a
number of stripes. We developed a microscopic picture of the
magnetic structure based on the results of our polarized neutron
studies which to some extent could be verified via Kerr microscopy.

\acknowledgments This study was supported by the DFG (SFB 491) and
by BMBF O3ZA6BC1 and by Forschungszentrum J\"ulich.

\end{document}